\documentclass[final,times,twocolumn]{elsarticle}
\usepackage{graphicx,amssymb,amsmath}
\usepackage{multirow,dcolumn,bm,latexsym,soul}
\newcolumntype{K}[1]{>{\centering\arraybackslash}p{#1}}

\journal{Physics Letters B}
\begin{document}
\newcommand{\be}{\begin{equation}}
\newcommand{\ee}{\end{equation}}
\newcommand{\bq}{\begin{eqnarray}}
\newcommand{\eq}{\end{eqnarray}}

\begin{frontmatter}

\title{Low-redshift constraints on homogeneous and isotropic universes with torsion}
\author[inst1]{C. M. J. Marques}\ead{cmd.marques@campus.fct.unl.pt}
\author[inst2,inst3]{C. J. A. P. Martins\corref{cor1}}\ead{Carlos.Martins@astro.up.pt}
\address[inst1]{Faculdade de Ci\^encias e Tecnologia, Universidade Nova de Lisboa, 2829-516 Caparica, Portugal}
\address[inst2]{Centro de Astrof\'{\i}sica da Universidade do Porto, Rua das Estrelas, 4150-762 Porto, Portugal}
\address[inst3]{Instituto de Astrof\'{\i}sica e Ci\^encias do Espa\c co, CAUP, Rua das Estrelas, 4150-762 Porto, Portugal}
\cortext[cor1]{Corresponding author}

\begin{abstract}
One of the possible extensions of Einstein's General Theory of Relativity consists in allowing for the presence of spacetime torsion. The form of the underlying torsion tensor can be chosen such that the homogeneity and isotropy of Friedmann-Lemaitre-Robertson-Walker universes is preserved, and it has been recently suggested that such universes may undergo accelerating phases. We use recent low-redshift data, coming from Type Ia Supernova and Hubble parameter measurements, to phenomenologically constrain this class of models under the so-called steady-state torsion assumption of a constant fractional contribution of torsion to the volume expansion. We start by considering models without a cosmological constant (where torsion itself would be expected to yield the current acceleration of the universe) finding, in agreement with other recent works, that these are strongly disfavoured by the data. We then treat these models as one-parameter extensions of $\Lambda$CDM, constraining the relative contribution of torsion to the level of a few percent in appropriate units. Finally, we briefly discuss how these constraints may be improved by forthcoming low-redshift data and check the robustness of our results by studying an alternative to the steady-state torsion parametrization.
\end{abstract}
\end{frontmatter}


\section{Introduction}
\label{sect1}

Understanding the physical mechanism behind the recent acceleration of the universe is among the most pressing tasks of observational cosmology. Whether this mechanism is a cosmological constant or a new dynamical degree of freedom---describing an additional fluid or a modification in the large-scale behaviour of gravity---is still unknown, and major observational efforts are in progress to address the issue \cite{Huterer}.

One of the natural extensions of Einstein's General Theory of Relativity consists in allowing for the presence of spacetime torsion. This was first considered almost one century ago by Cartan \cite{Cartan}, and had a brief revival of interest in the 1960s due to seminal works by Kibble and Sciama \cite{Kibble,Sciama}, but its observational consequences are relatively unexplored. In the modern cosmological context, it is interesting to note that the form of underlying torsion tensor can be chosen such that the homogeneity and isotropy of Friedmann-Lemaitre-Robertson-Walker universes is preserved \cite{Tsamparlis}. This scenario has been recently reconsidered \cite{Torsion1} and it has been noted that such universes may undergo accelerating phases \cite{Torsion2}.

Here we use recent low-redshift data, coming from the Pantheon Type Ia Supernova compilation by Riess {\it et al.} \cite{Riess} and the compilation of 38 Hubble parameter measurements by Farooq {\it et al.} \cite{Farooq} to constrain this class of models. We start by considering models without a cosmological constant (where torsion would be expected to yield the current acceleration of the universe) finding that these are strongly disfavoured by the data. To the extent that a comparison is possible, our results here are in agreement with other recent work \cite{Pereira}, which uses different assumptions and data (as we discuss below). We then treat these models under a more general assumption that they are one-parameter extensions of $\Lambda$CDM, constraining the relative contribution of torsion to the level of a few percent (in appropriate units, also defined below). We also briefly discuss how these constraints may be improved by forthcoming low-redshift data, specifically from a combination of next generation supernova measurements from WFIRST and novel measurements of the redshift drift of cosmological objects following the Hubble flow from the ELT.

The plan of the rest of the work is as follows. We start by briefly describing the derivation of the Einstein and continuity equations for these models in Sect. \ref{sect2}, and also comment on the steady-state torsion parametrization which we will use in most of what follows. Constraints on these models from the aforementioned current data are presented in Sect. \ref{sect3}, while in Sect. \ref{sect4} we look beyond the current status and present forecasts for analogous constraints from next-generation experiments. In Sect. \ref{sect5} we briefly study and constrain an alternative to the steady-state torsion parametrization (in order to test how generic are the results obtained with our primary choice of parametrization), and finally we present some conclusions in Sect. \ref{sect6}


\section{Homogeneous and isotropic models with torsion}
\label{sect2}

Spatially homogeneous and isotropic cosmologies including torsion have recently been studied in \cite{Torsion1,Torsion2}. In theories with torsion there is a further degree of freedom (in addition to the usual metric), which also gravitates. While there is at present no experimental or observational evidence for the presence of this degree of freedom, it is important to study its possible effects, and here we do this for the recent universe. The torsion field can be chosen such as to preserve the homogeneity and isotropy of Friedmann-Lemaitre-Robertson-Walker models \cite{Tsamparlis}, and as recently discussed in \cite{Torsion1,Torsion2} torsion can play a non-trivial cosmological role and in some cases lead to accelerating universes. In what follows we briefly summarize the results of these theoretical works, leading to the Friedmann, Raychaudhuri and continuity equations for these models, which will be the starting point of our observational analysis.

Mathematically, the torsion tensor $S^\xi_{\mu\nu}$ is defined as the antisymmetric part of the affine connection,
\be
S^\xi_{\mu\nu}=\Gamma^\xi_{[\mu\nu]}\,;
\ee
the symmetric part of the connection are the usual Christoffel symbols. Physically, this defines relation between the intrinsic angular momentum (i.e., the spin) of matter with the geometric properties of the underlying spacetime. The only non-trivial contraction of the torsion tensor is the torsion vector, $S_\mu=S^\nu_{\mu\nu}$. The general field equations including torsion are known as the Einstein-Cartan equations. Nominally the Einstein equations retain the usual form
\be
R_{\mu\nu}-\frac{1}{2}Rg_{\mu\nu}+\Lambda g_{\mu\nu} =\kappa T_{\mu\nu}\,,
\ee
where $g_{\mu\nu}$ is the metric, $R_{\mu\nu}$ and $R$ are the Ricci tensor and scalar, $T_{\mu\nu}$ is the energy-momentum tensor, $\Lambda$ is Einstein's cosmological constant and $\kappa=8\pi G$. However, note that the presence of torsion implies that the Ricci tensor and the energy-momentum tensor are not symmetric. The Cartan equations relate the torsion tensor to the spin tensor, denoted $s_{\mu\nu\xi}$, as follows
\be
S_{\mu\nu\xi}=-\frac{\kappa}{4}\left(2s_{\nu\xi\mu}+g_{\xi\mu}s_\nu-g_{\mu\nu}s\xi \right)\,,
\ee
and similarly for the torsion and spin vectors one has $S_\mu=-\kappa s_\mu/4$.

Generically speaking, models containing torsion do not lead to homogeneous and isotropic universes. However, as discussed in \cite{Tsamparlis}, there is a particular choice of torsion tensor which does lead to such universes. It is convenient to define it by considering a $3+1$ spacetime decomposition, introducing a timelike 4-velocity field $u_\mu$ (satisfying $u_\mu u^\mu=-1$) and using it to decompose the metric into $g_{\mu\nu}=h_{\mu\nu}-u_\mu u_\nu$, with the tensor $h_{\mu\nu}$ being symmetric and orthogonal to the vector $u_\mu$. With these definitions, the required torsion tensor has the form
\be
S_{\mu\nu\xi}=2\phi(t) h_{\mu[\nu}u_{\xi]}\,;
\ee
the scalar function $\phi$ must depend only on time (a spatial dependence would violate the homogeneity assumption), but is otherwise arbitrary.

Making the standard assumption of treating the metric and the torsion as independent objects, the line element of a FRW-like spacetime with torsion should be the same as in the standard Riemann case, namely
\be
d{\cal s}^2=-dt^2+a^2\left[\frac{dr^2}{1-Kr^2}+r^2d\Omega^2\right]\,,
\ee
where $a$ is the scale factor and $K$ is the 3-curvature. Standard computational techniques then lead to the following Friedmann, Raychaudhuri and continuity equations \cite{Torsion1,Torsion2}
\be
H^2=\frac{1}{3}\kappa\rho-\frac{K}{a^2}+\frac{1}{3}\Lambda-4\phi^2-4H\phi
\ee
\be
\frac{\ddot a}{a}=-\frac{1}{6}(\rho+3p)+\frac{1}{3}\Lambda-2{\dot\phi}-2H\phi
\ee
\be
{\dot\rho}=-3H\left(1+2\frac{\phi}{H}\right)(\rho+p)+4\phi\left(\rho+\frac{\Lambda}{\kappa}\right)\,.
\ee
Here the dot denotes a derivative with respect to physical time, $H={\dot a}/a$ is the Hubble parameter, and finally $\rho$ and $p$ are the density and pressure. In what follows we will assume $K=0$ and barotropic fluids with a constant equation of state $p=w\rho$.

We can also define the usual present-day fractions of matter and dark energy, $\Omega_m=\kappa\rho_{0}/3H_0^2$ and $\Omega_\Lambda=\Lambda/3H_0^2$, and we can similarly define a torsion contribution
\be\label{torphi}
\Omega_\phi=-4\left(\frac{\phi_0}{H_0}\right)\left[1+\frac{\phi_0}{H_0}\right]\,.
\ee
In most of what follows we will make the usual simplifying assumption (also adopted in \cite{Torsion1,Pereira}, where it is dubbed steady-state torsion) that the relative torsion contribution to the expansion remains constant in time,
\be
\frac{\phi}{H}=\lambda=const.\,,
\ee
and we will usually express observational constraints directly in terms of the model parameter $\lambda$. The exception is Sect. \ref{sect5}, where we will briefly consider an alternative phenomenological parametrization for the torsion field $\phi$.

The analysis of \cite{Torsion1} leads to a bound from primordial nucleosynthesis of $-0.01\lesssim\lambda\lesssim0.02$ (in the absence of an explicit statement therein, we assume that this is a one-sigma bound). The more recent analysis of \cite{Pereira} uses models with up to 4 free parameters, assuming $\Lambda=0$ and $w=0$ throughout, but not assuming flat universes. Under those assumptions they find that scenarios with $\phi=const$ or $\phi\propto H$ are observationally disfavoured (we will confirm the latter below), and toy models with 3 or 4 free parameters would be needed in order to have realistic values of the matter density and Hubble constant (though such models will have strong degeneracies between some of these parameters). Our analysis differs in focusing on models with fewer and/or better motivated parameters: we will assume flat universes, but allow for non-zero values of the cosmological constant and of the matter equation of state.

For numerical purposes, in the following sections it is convenient to define a dimensionless density ($r$) via $\rho=r\rho_0$, as well as the dimensionless Hubble parameter $E=H/H_0$, and write the Friedmann and continuity equations as a function of redshift as follows
\be
E^2(z)=1+\frac{r(z)-1}{(1+2\lambda)^2}\Omega_m
\ee
\be
(1+z)\frac{dr}{dz}=3(1+w)r+2\lambda\left[2+(1+3w)r-\frac{2}{\Omega_m}(1+2\lambda)^2\right]\,,
\ee
where we have made use of the flatness condition (leading to a relation between $\lambda$, $\Omega_m$ and $\Omega_\Lambda$ which allows us to eliminate the latter) and are also assuming that $\lambda\neq-1/2$ and $\Omega_m\neq0$. We will deal with the particularly simple case $\Omega_\Lambda=0$ separately at the beginning of the next section.


\section{Constraints from current data}
\label{sect3}

We now use recent low-redshift background cosmology data to constrain these models, under various different assumptions about the underlying parameter space. Specifically, we use two main data sets. The first is the compressed data from the Pantheon compilation \cite{Riess}. We note that the values reported in the arXiv and the published versions are slightly different; in what follows we used the values from the published version. The 1049 supernova measurements in the range $0<z<2.3$ are compressed into 6 correlated measurements of $E^{-1}(z)$ (where $E(z)$is again the dimensionless Hubble parameter) in the redshift range $0.07<z<1.5$. It has been shown \cite{Riess} that this compressed data set provides a nearly identical characterization of dark energy as the full supernova sample, thus making it an efficient compression of the raw data. Our second data set is the recent heterogeneous compilation of 38 Hubble parameter measurements in the redshift range  $0.07<z<2.36$ \cite{Farooq}.

We carried out a standard likelihood analysis with up to three different free parameters: $\Omega_m$, $\lambda$ and the equation of state $w$. We always assume uniform priors on $\lambda$ and $w$. For the matter density we will in some cases use a Planck-like prior, $\Omega_m=0.315\pm0.007$ \cite{Planck}, otherwise the prior is also uniform. Another relevant parameter is the Hubble constant, but this is always analytically marginalized in our analysis, following the procedure detailed in \cite{Basilakos}.

\subsection{No cosmological constant}

\begin{figure*}
\begin{center}
\includegraphics[width=3.2in,keepaspectratio]{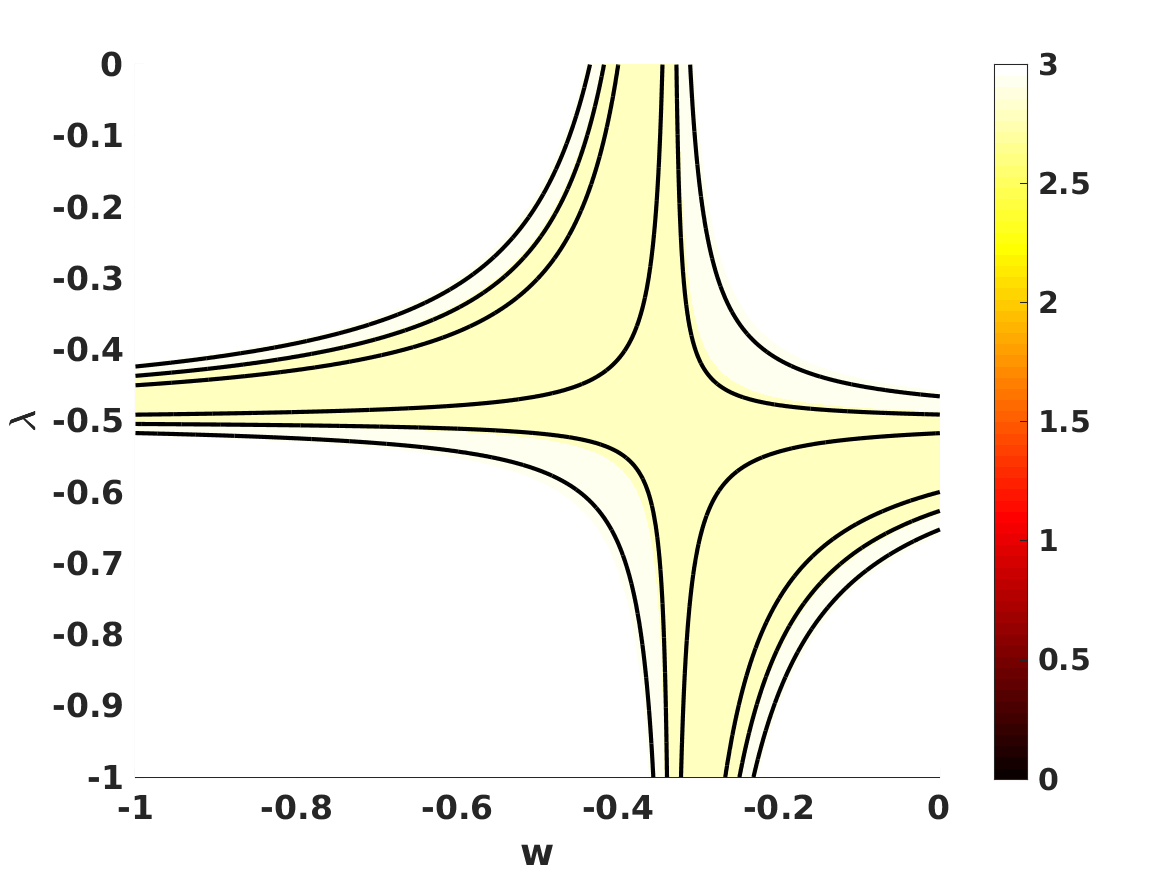}
\includegraphics[width=3.2in,keepaspectratio]{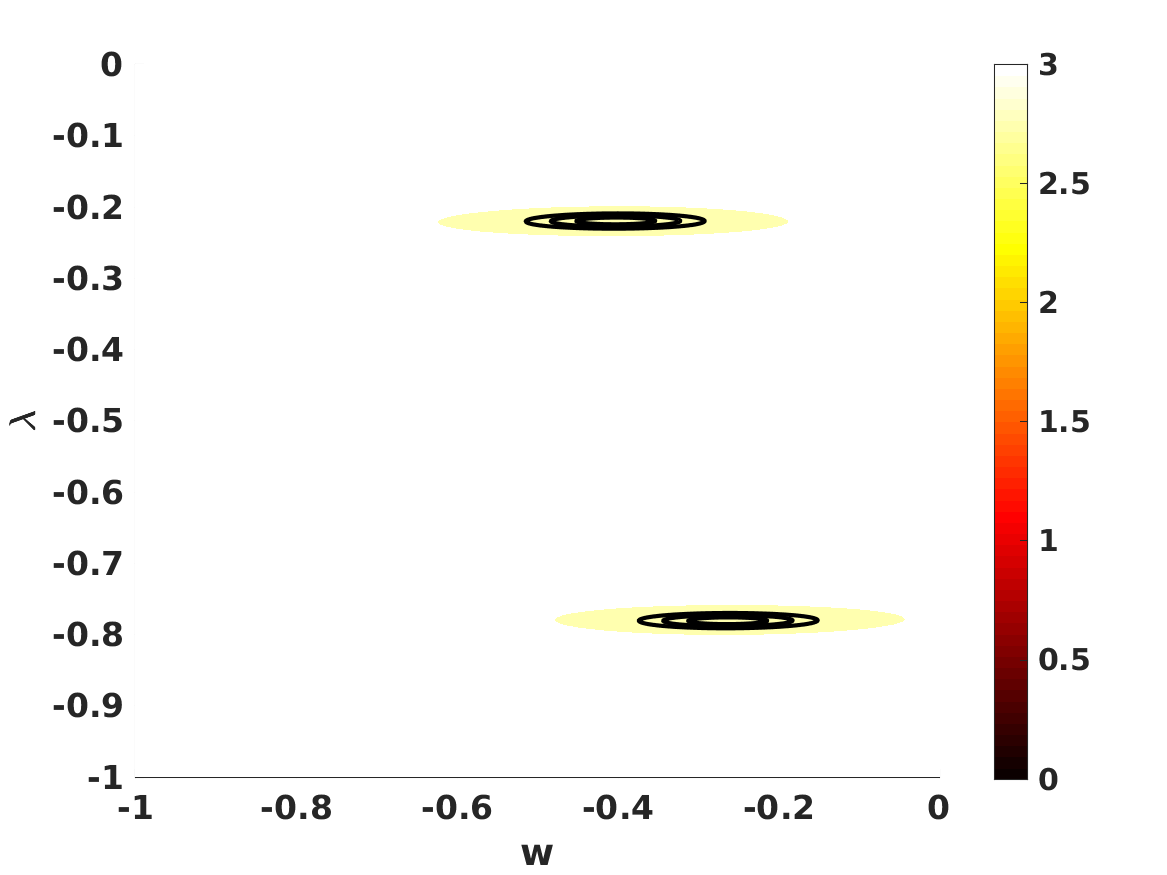}
\end{center}
\caption{\label{fig1}Constraints on the $\lambda$--$w$ parameter space for $\Omega_\Lambda=0$. Left and right panels are respectively for the cases without and with the Planck prior. The black lines represent the one, two and three sigma confidence levels, and the colormap depicts the reduced chi-square of the fit: for the best-fit choice of model parameters, this is respectively 2.78 and 2.71, confirming that this is a very poor fit to the data.}
\end{figure*}

We start with the simple case $\Omega_\Lambda=0$. In this case the dimensionless Friedmann equation has the analytic solution
\be
E^2(z)=(1+z)^{3(1+w)+2\lambda(1+3w)}\,;
\ee
note that the matter density does not explicitly appear because it and $\lambda$ are not independent, being related by
\be
\Omega_m=(1+2\lambda)^2\,.
\ee
Clearly there will be a degeneracy between $\lambda$ and $w$, although it can be broken by a prior on the matter density---for which we will use Planck, as mentioned above. Figure \ref{fig1} summarizes the results of our analysis, without and with this prior. Note that in this and subsequent figures the pairs of panels are always displayed with identical axis ranges, to facilitate the comparison and highlight the impact of the prior.

The most striking result is that this is a very poor fit to the data: the reduced chi-squares for the best fit choice of parameters are respectively 2.78 without the prior and 2.71 with the prior. In the latter case the relation
\be
\lambda=\frac{1}{2}(-1\pm\sqrt{\Omega_m})\,
\ee
determines the two possible values for the torsion ratio, $\lambda=-0.22$ or $\lambda=-0.78$; these would correspond to $\Omega_\phi=1-\Omega_m\sim0.685$. The preferred values for the equation of state $w$ are -0.39 and -0.28 without the prior and -0.40 and -0.26 with the prior, respectively.

Having shown that these models can't be fundamental alternatives to $\Lambda$CDM, in which torsion would be responsible for the recent acceleration of the universe, we move on to consider the more phenomenological scenario where torsion is an extension of standard $\Lambda$CDM cosmology, enabling constraints on the observationally allowed amount of torsion.

\subsection{Cosmological constant with $w=0$}

We now return to the generic case where $\Omega_\Lambda$ need not vanish, so we are effectively treating these models as one-parameter extensions of $\Lambda$CDM, to which model they reduce when $\lambda=0$. We start by assuming that matter has the standard equation of state, $w=0$. As in the previous subsection will separately consider the cases without and with the aforementioned Planck prior on the matter density.

\begin{figure*}
\begin{center}
\includegraphics[width=3.2in,keepaspectratio]{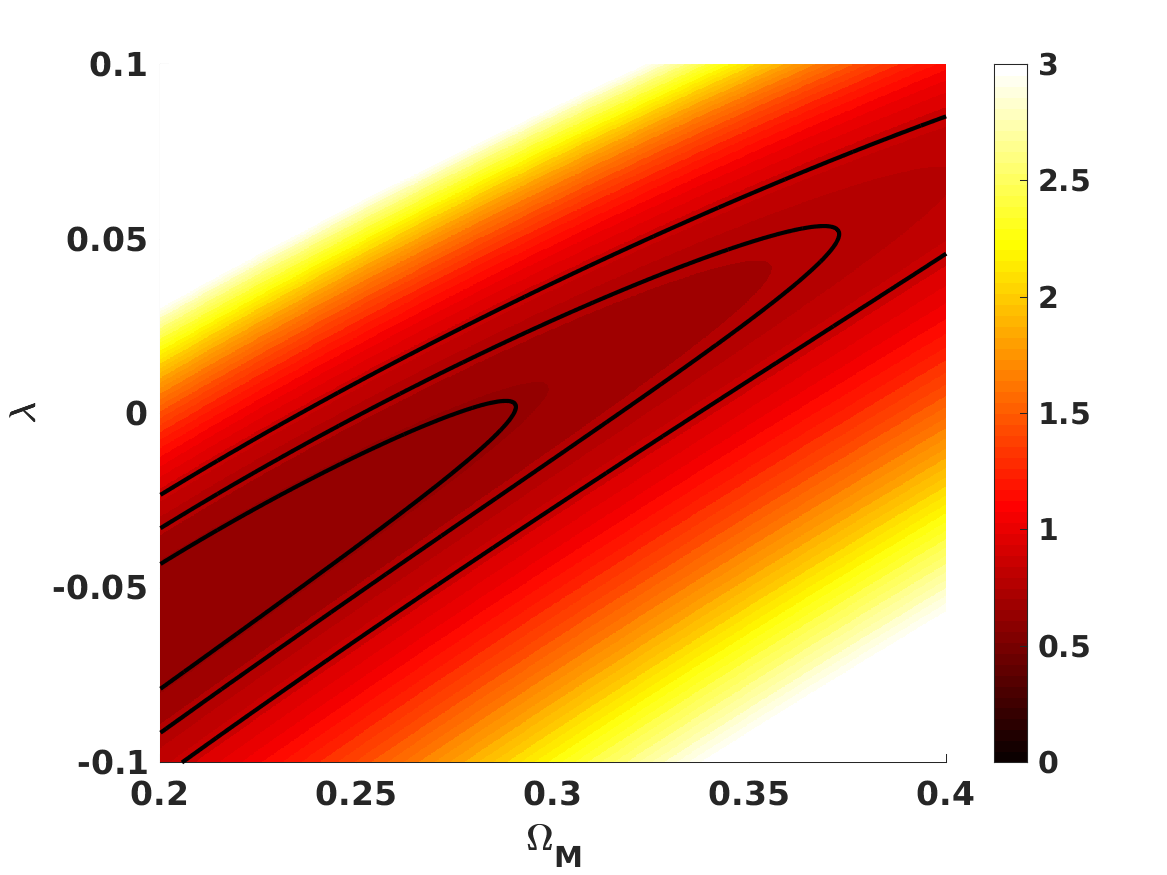}
\includegraphics[width=3.2in,keepaspectratio]{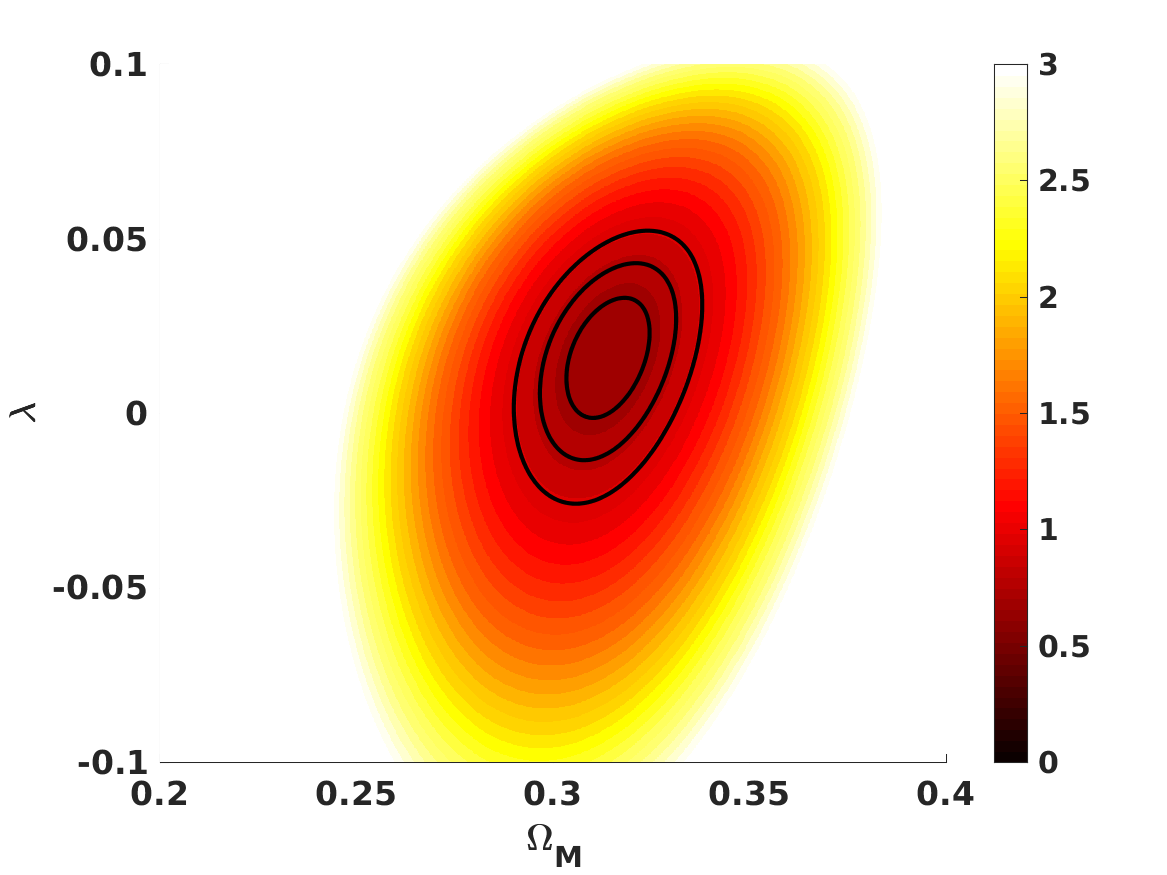}
\end{center}
\caption{\label{fig2}Constraints on the $\lambda$--$\Omega_m$ parameter space for $w=0$.  Left and right panels are respectively for the cases without and with the Planck prior. The black lines represent the one, two and three sigma confidence levels, and the colormap depicts the reduced chi-square of the fit.}
\end{figure*}

Figure \ref{fig2} summarizes the results of this analysis. Starting with the case without the Planck prior, we find the following one-sigma posterior likelihoods for the two free parameters
\be
\lambda_{w=0}=-0.07^{+0.05}_{-0.04}\,,
\ee
\be
\Omega_{m,w=0}=0.18^{+0.06}_{-0.03}\,;
\ee
there is a clear degeneracy between the two parameters, and there is a one-sigma preference for a non-zero (negative) $\lambda$, with a correspondingly smaller value of the matter density. However, at the two-sigma level the results are consistent with $\Lambda$CDM. The inclusion of the Planck prior breaks the degeneracy and significantly improves the constraints. Additionally (and unsurprisingly) the Planck prior makes the results more consistent with $\Lambda$CDM; the one-sigma posterior likelihood for the torsion parameter is found to be
\be
\lambda_{(w=0, Planck)}=0.02^{+0.01}_{-0.02}\,,
\ee 
which is consistent with the null result at just over one sigma. The Planck prior shifts the preferred value of the matter density to significantly larger values, and the positive correlation between the two parameters makes the preferred value of $\lambda$ correspondingly larger. For the best fit value, and according to Eq. \ref{torphi}, the torsion contribution would therefore be $\Omega_\phi\sim-0.07$.

\subsection{Cosmological constant with $w\neq0$}

We can extend the results of the previous subsection by allowing for a non-zero (but still constant) equation of state. There is a weak degeneracy between $w$ and the other model parameters, so although the constraints become weaker (as they must), both parameters are still well constrained by the data. On the other hand, without the Planck prior the matter density is not constrained, even at one sigma.

\begin{figure*}
\begin{center}
\includegraphics[width=3.2in,keepaspectratio]{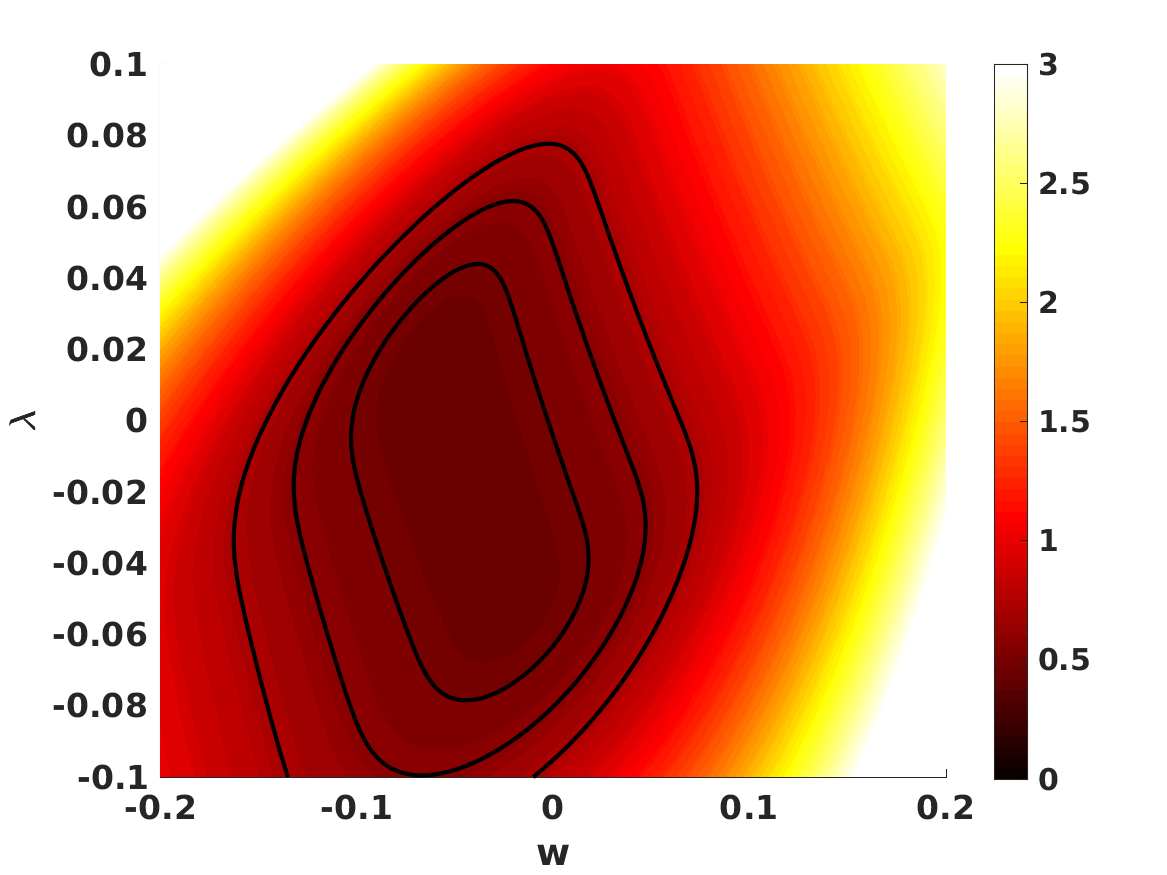}
\includegraphics[width=3.2in,keepaspectratio]{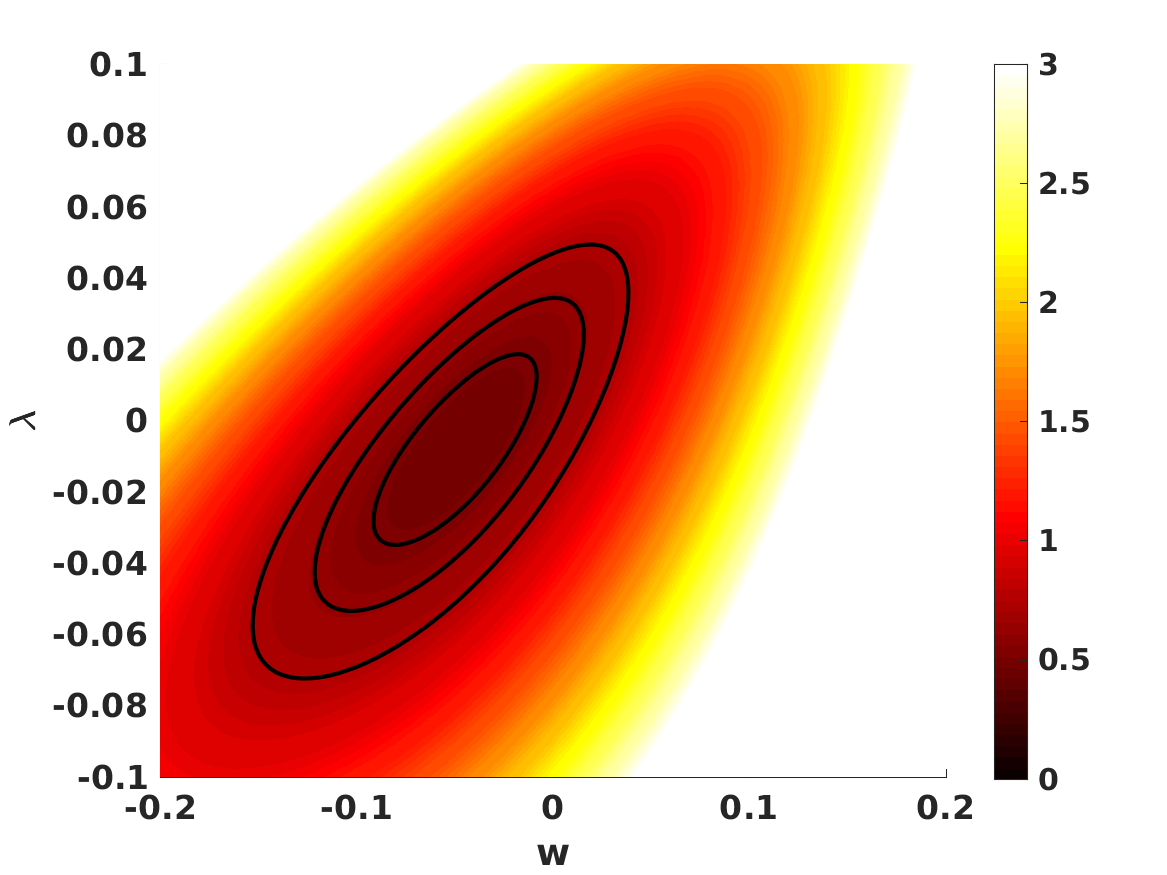}
\includegraphics[width=3.2in,keepaspectratio]{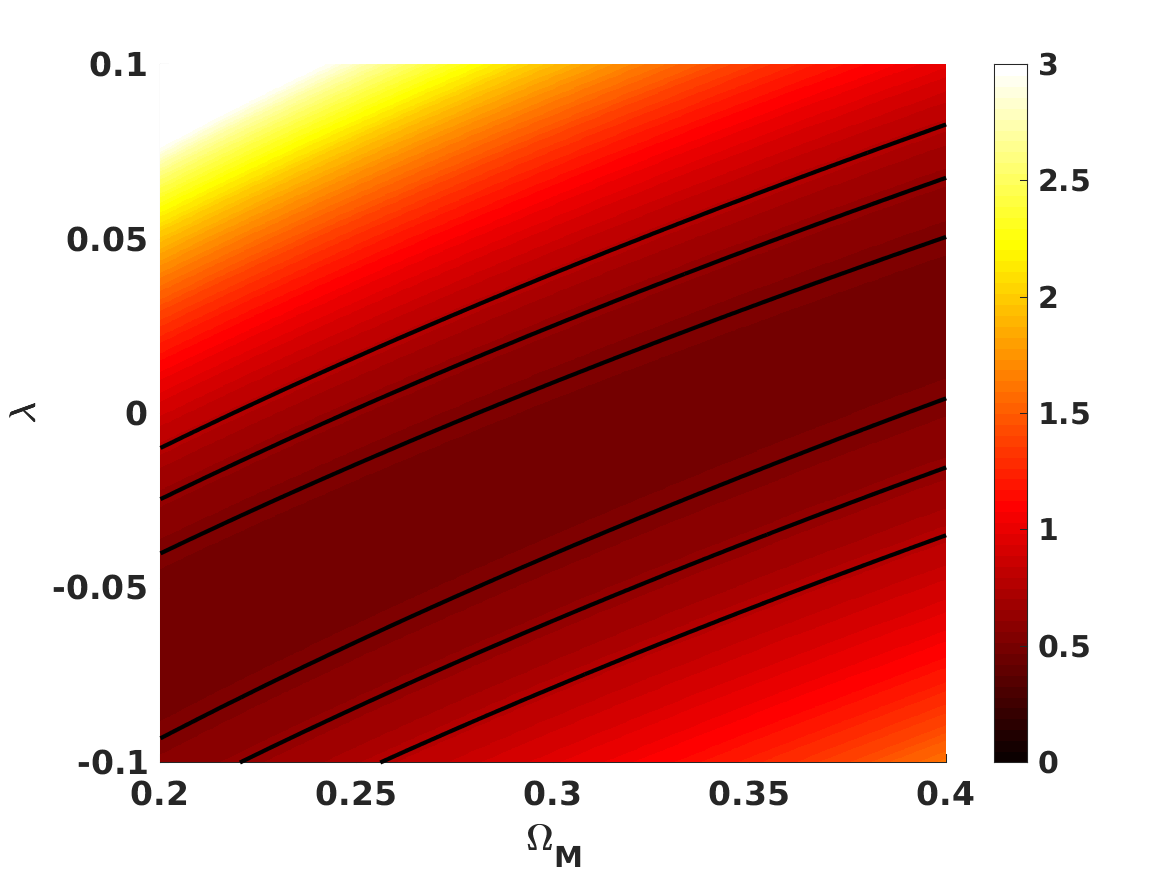}
\includegraphics[width=3.2in,keepaspectratio]{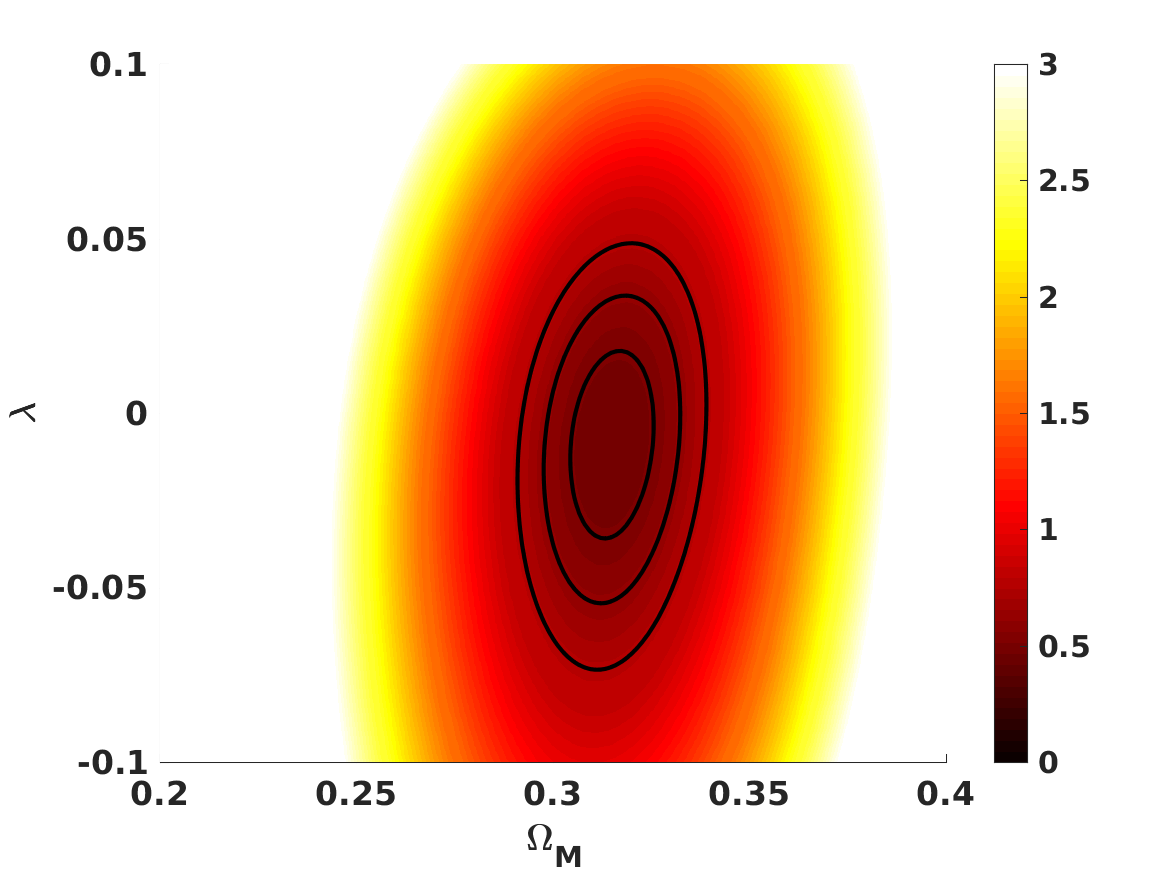}
\includegraphics[width=3.2in,keepaspectratio]{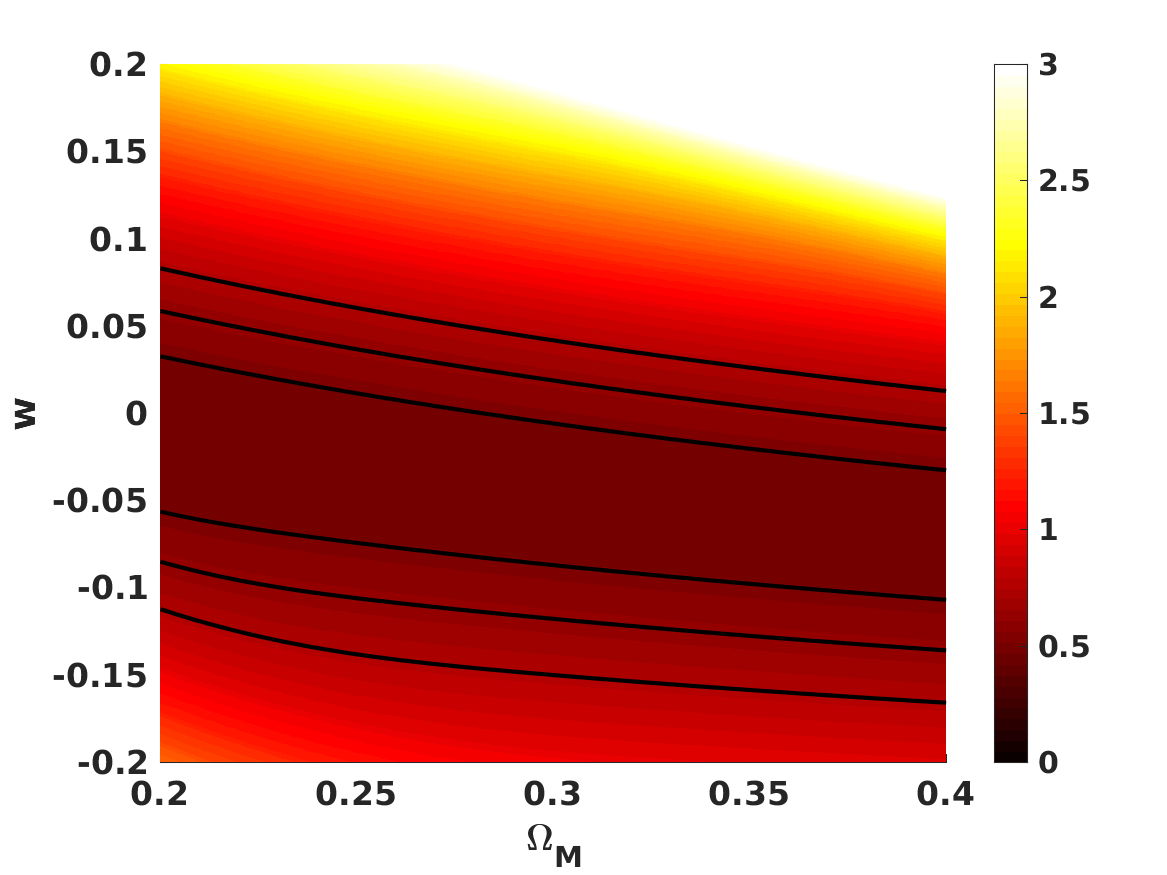}
\includegraphics[width=3.2in,keepaspectratio]{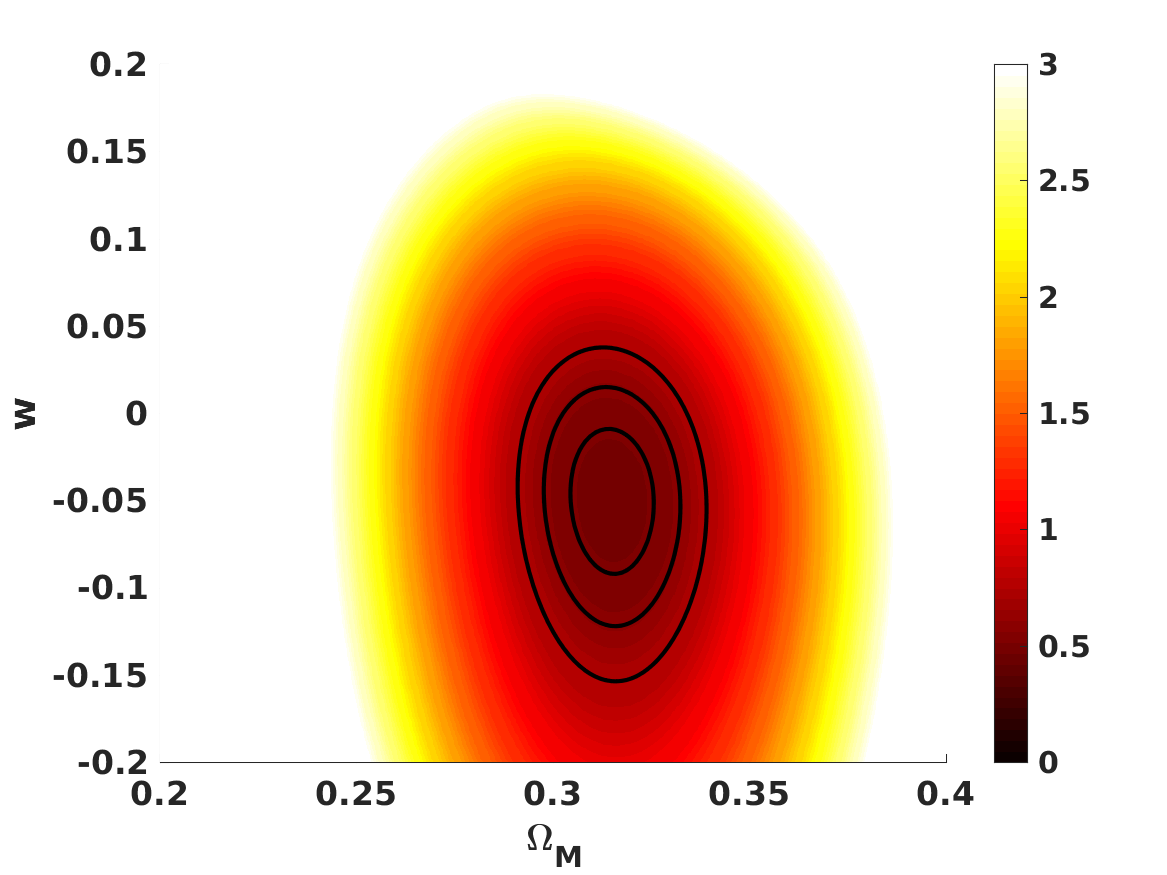}
\end{center}
\caption{\label{fig3}Likelihood constraints in the various 2D planes of the $\lambda$--$w$--$\Omega_m$ parameter space.  Left and right side panels are respectively for the cases without and with the Planck prior. The black lines represent the one, two and three sigma confidence levels, and the colormap depicts the reduced chi-square of the fit.}
\end{figure*}

Figure \ref{fig3} summarizes the results of this analysis. Starting with the case without the Planck prior, the one-sigma posterior likelihood for the torsion parameter is
\be
\lambda_{(w\neq0)}=-0.02\pm0.04\,,
\ee
while for the equation of state we have
\be
w=-0.05\pm0.04\,;
\ee
there is therefore a small (statistically not significant) preference for a negative equation of state for matter. Moreover, we see that this equation of state is quite tightly constrained, which explains why the constraint on $\lambda$ is not significantly changed with respect to the $w=0$ case. Indeed the main difference in allowing for a non-zero equation of state is that there is no longer any constraining power on the matter density. This can again be circumvented by adding the Planck prior, which leads to
\be
\lambda_{(w\neq0, Planck)}=-0.01\pm0.02\,;
\ee
compared to the $w=0$ case the best-fit value has changed sign, and the constraint is now consistent with the null result at one sigma. When compared to the $w\neq0$ constraint without the matter prior the sensitivity is improved by a factor of two. For the best fit value the torsion contribution would be $\Omega_\phi\sim0.03$. On the other hand, the constraint on the matter equation of state is relatively less affected
\be
w_{Planck}=-0.05\pm0.03\,;
\ee
this is consistent with the standard result ($w=0$) at the two sigma confidence level.

\section{Forecasts of future constraints}
\label{sect4}

Here we briefly discuss how the constraints discussed in the previous section might be improved by future observations. Specifically, we consider measurements of the redshift drift by the Extremely Large Telescope (ELT) \cite{Liske,Alves}, which will directly probe the expansion of the universe in the deep matter era, as well as an improved supernova data set.

The redshift drift of an astrophysical object following the Hubble flow is given by \cite{Sandage,Liske}
\be
\Delta z=\tau_{obs} H_0 \left[1+z-E(z)\right]\,,
\ee
where $\tau_{obs}$ is the observation time span, although the actual observable is a spectroscopic velocity
\be
\Delta v=k\tau_{obs}h\left[1-\frac{E(z)}{1+z}\right]\,,
\ee
which we expressed in terms of $E(z)$ and $h=H_0/(100 km/s/Mpc)$; we also introduced the normalization constant $k$, which has the value $k=3.064$ cm/s if $\tau_{obs}$ is expressed in years. The uncertainty in the velocity measurement for the ELT is expected to be \cite{Liske,HIRES}
\be
\sigma_v(z)=1.35\left(\frac{2370}{S/N}\right)\sqrt{\frac{30}{N_{qso}}}\left(\frac{5}{1+z_{qso}}\right)^\lambda\,,
\ee
where $S/N$ denotes the signal to noise of the spectra available at the redshift bin $z_{qso}$ and $N_{qso}$ is the number of quasars observed at that redshift. The exponent of the last term is $\lambda=1.7$ for $z_{qso}\le4$ and $\lambda=0.9$ for $z_{qso}>4$. We assume a realistic observation program with a time span of $\tau_{obs}=20$ years, a signal to noise $S/N=3000$ in each measurement, and three different measurements at redshift bins centered at $z_{qso}=2.5, 3.5, 5.0$, each based on combining spectra from sets of $N_{qso}=10$ quasars \cite{HIRES}. 

The work leading to the Pantheon compilation \cite{Riess} also discusses a future data set of supernova measurements from the proposed WFIRST satellite \cite{WFIRST}. Their analysis leads to the following values for percent errors on $E(z)$:  $\sigma=1.3, 1.1, 1.5, 1.5, 2.0, 2.3, 2.6, 3.4, 8.9$ for the nine redshift bins centered at $z=0.07, 0.20, 0.35, 0.60, 0.80, 1.00, 1.30, 1.70, 2.50$, respectively. These measurements are not fully independent, as there are pairwise correlations among some of the measurements, but the authors state that these effects are small, so in the absence of publicly available information on them we ignore them in the analysis that follows. Their simulated data set was also obtained under the assumption of a flat universe.

\begin{figure}
\begin{center}
\includegraphics[width=3.2in,keepaspectratio]{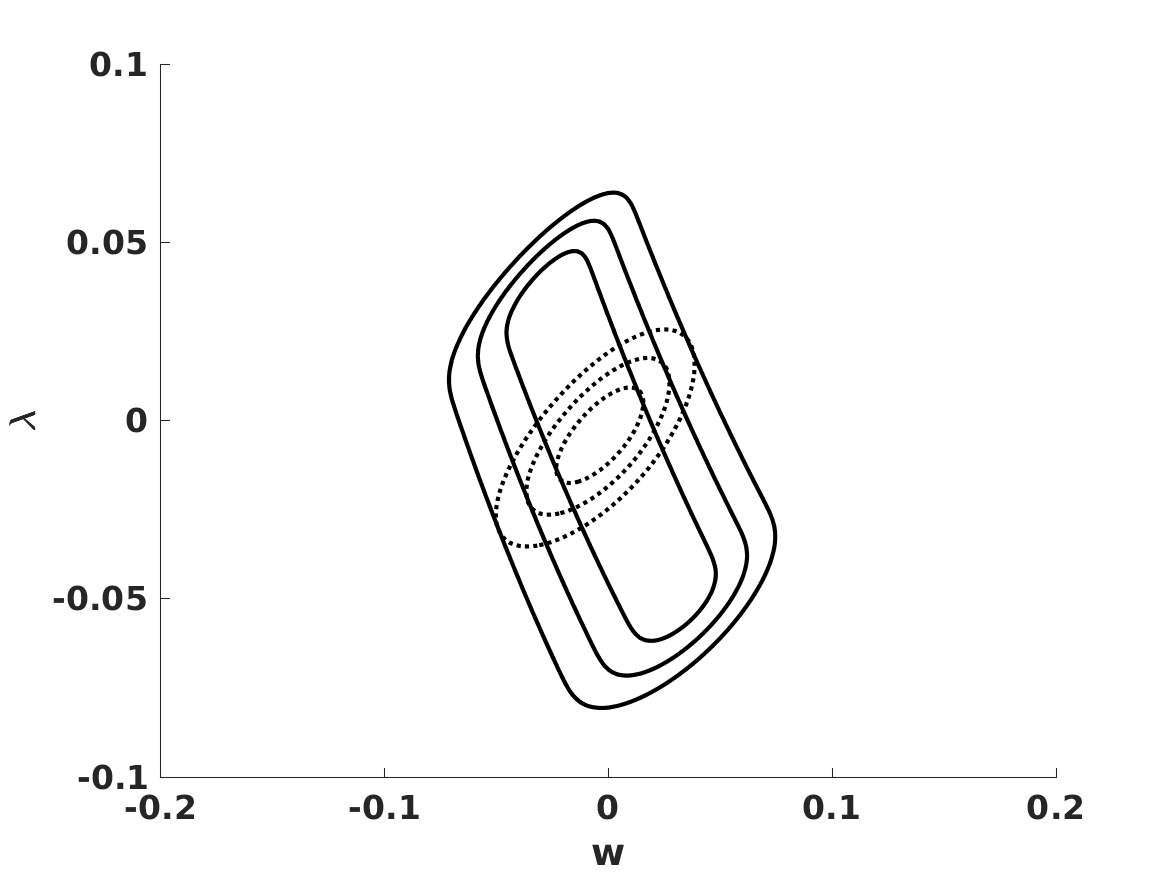}
\includegraphics[width=3.2in,keepaspectratio]{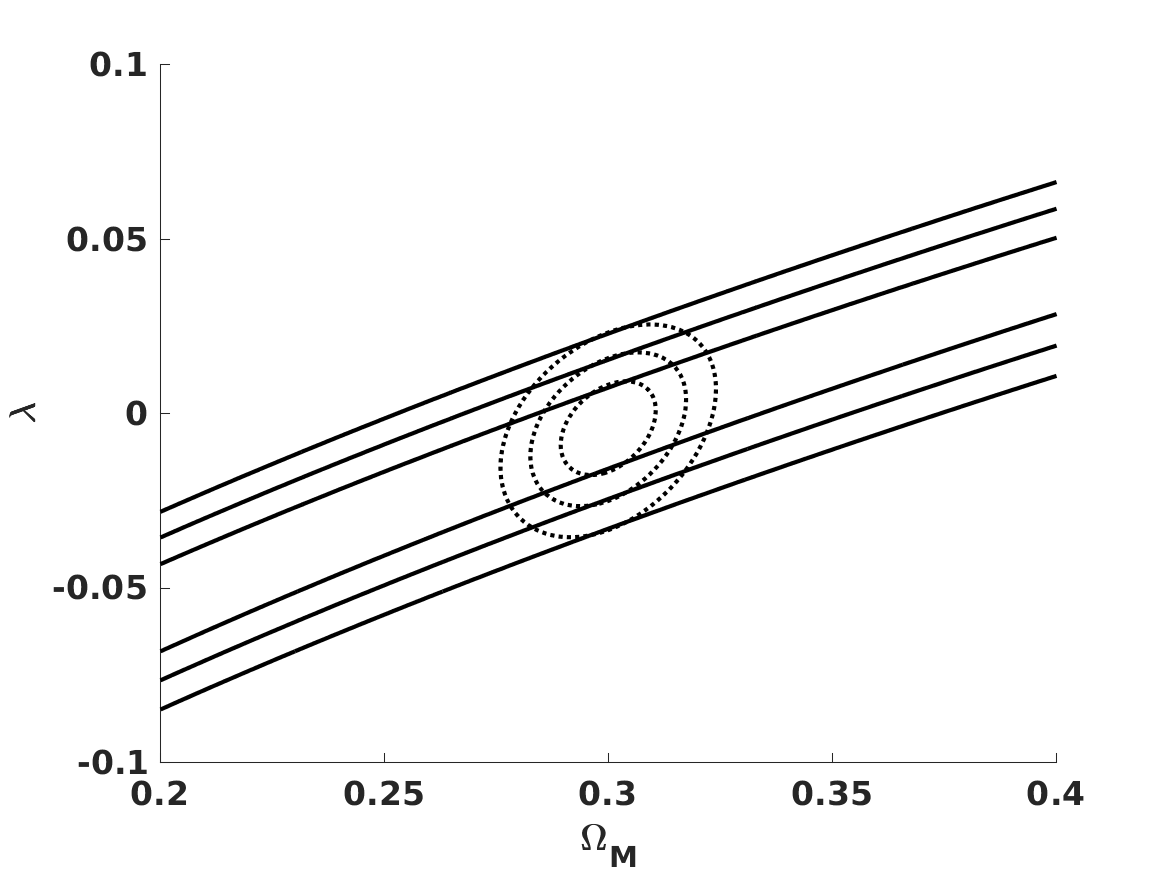}
\includegraphics[width=3.2in,keepaspectratio]{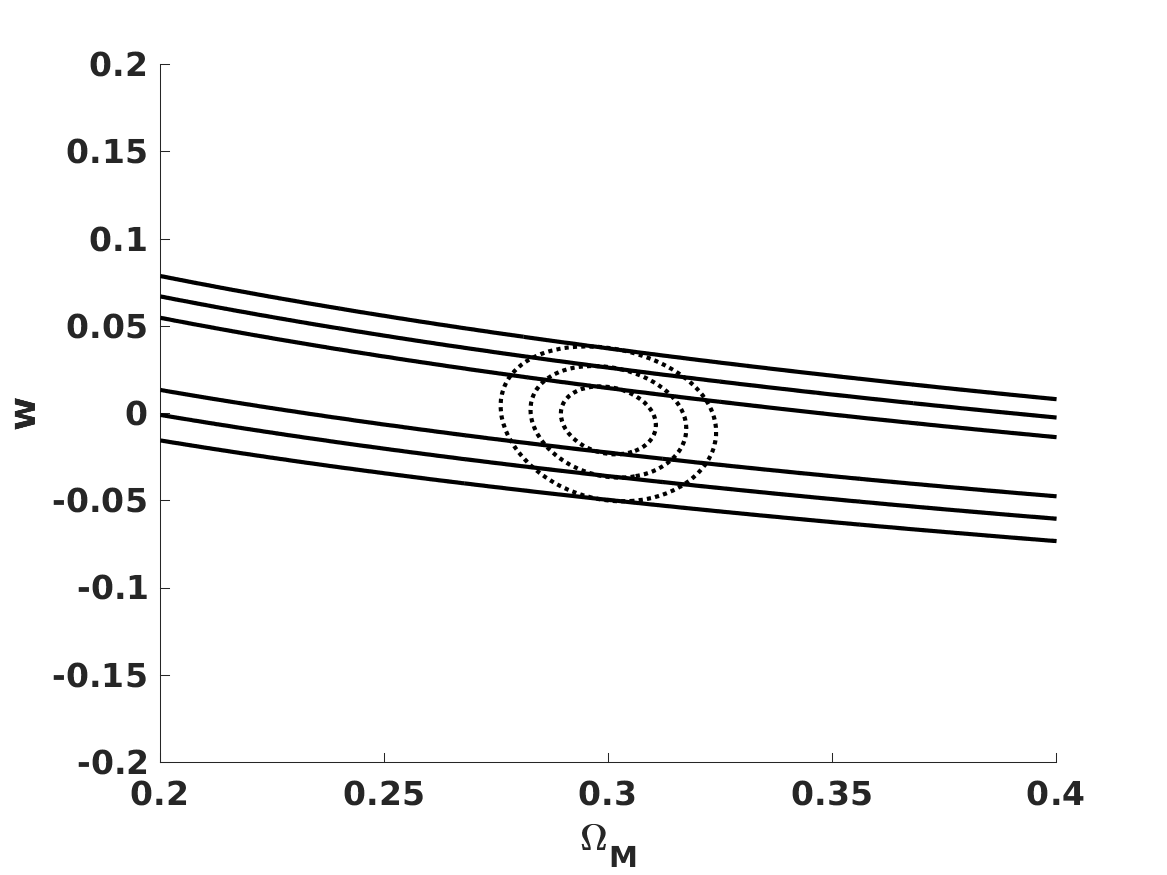}
\end{center}
\caption{\label{fig4}Likelihood constraints in the various 2D planes of the $\lambda$--$w$--$\Omega_m$ parameter space for the future (simulated) data sets described in the main text. Dashed and solid lines respectively depict the cases with and without priors. One, two and three sigma confidence levels are represented in both cases.}
\end{figure}

We thus forecast constraints on the models under consideration for a combined mock data set of ELT redshift drift and WFIRST supernova measurements, both on their own and complemented by a Planck-like prior that has an uncertainty on the matter density $\sigma(\Omega_m)=0.007$.  We assume a fiducial $\Lambda$CDM model with $\Omega_m=0.3$, $\lambda=0$, $w=0$ and $h=0.7$.

The results are summarized in Fig. \ref{fig4}. Similarly to the results for current data in the previous section, we again note that the matter density is unconstrained without a matter prior. Our analysis indicates that constraints on $w$ without the Planck prior are comparable to current constraints with the prior, while those on $\lambda$ are weaker by about a factor of three. On the other hand, with a Planck-level prior on the matter density, we forecast the following one-sigma uncertainties
\be
\sigma_\lambda=0.009
\ee
\be
\sigma_w=0.013\,,
\ee
which improve on the current ones (discussed in the previous section) by about a factor of two. The former corresponds to a one-sigma upper bound on the torsion contribution of $\Omega_\phi<0.036$.


\section{An alternative parametrization}
\label{sect5}

Our analysis so far has relied on the steady-state torsion assumption for the torsion field, $\phi=\lambda H$, which has also been used in the published literature. In this section we test the robustness of our analysis by briefly considering an alternative assumption.

From the phenomenological and dimensional analysis points of view, the most natural alternative parametrization to making the torsion field proportional to the Hubble parameter (i.e., the square root of the left-hand side of the Friedmann equation) is to make the it proportional to the square root of the matter density (i.e., the square root of the right-hand side of the Friedmann equation). Specifically, we now assume
\be
\phi=\frac{1}{2}\epsilon\sqrt{\frac{\kappa\rho}{3}}\,,
\ee
where $\epsilon$ is the free parameter replacing $\lambda$. In this case, and keeping the rest of our previously stated assumptions, the Friedmann and continuity equations now have the following form (written in terms of the dimensionless versions of the Hubble parameter and density)
\be
E(z)=\sqrt{\Omega_mr(z)+\Omega_\Lambda}-\epsilon\sqrt{\Omega_mr(z)}\,,
\ee
where the assumption of zero spatial curvature requires $\Omega_m+\Omega_\Lambda=(1+\epsilon\sqrt{\Omega_m})^2$, and
\be
(1+z)\frac{dr}{dz}=3(1+w)r+\epsilon\frac{\sqrt{\Omega_mr}}{E(z)}\left[3(1+w)r-2\left(r+\frac{\Omega_\Lambda}{\Omega_m}\right)\right]\,.
\ee

We now repeat the analysis of the previous section for this parametrization. Firstly, in the $\Omega_\Lambda=0$ case and with additional assumption of a small $\epsilon$, we have the following solution
\be
E^2(z)=(1+z)^{3(1+w)+\epsilon(1+3w)}\,,
\ee
with the matter density and $\epsilon$ being related via $\sqrt{\Omega_m}(1-\epsilon)=1$; this is therefore akin to the $\lambda$ case, and it is easy to see that it provides an equally poor fit to the data. Indeed, in the perturbative limit of small $\lambda$ and $\epsilon$ one simply has $\epsilon=2\lambda$, which is to be expected since in the absence of both a spatial curvature and a cosmological constant the Friedmann equation implies that Hubble parameter and matter density are approximately proportional to each other (with the torsion field providing a small correction to this behaviour). However, note that this simple relation between the two parameters is only a rough approximation in the general case with a cosmological constant and/or with non-negligible amounts of torsion.

\begin{figure}
\begin{center}
\includegraphics[width=3.2in,keepaspectratio]{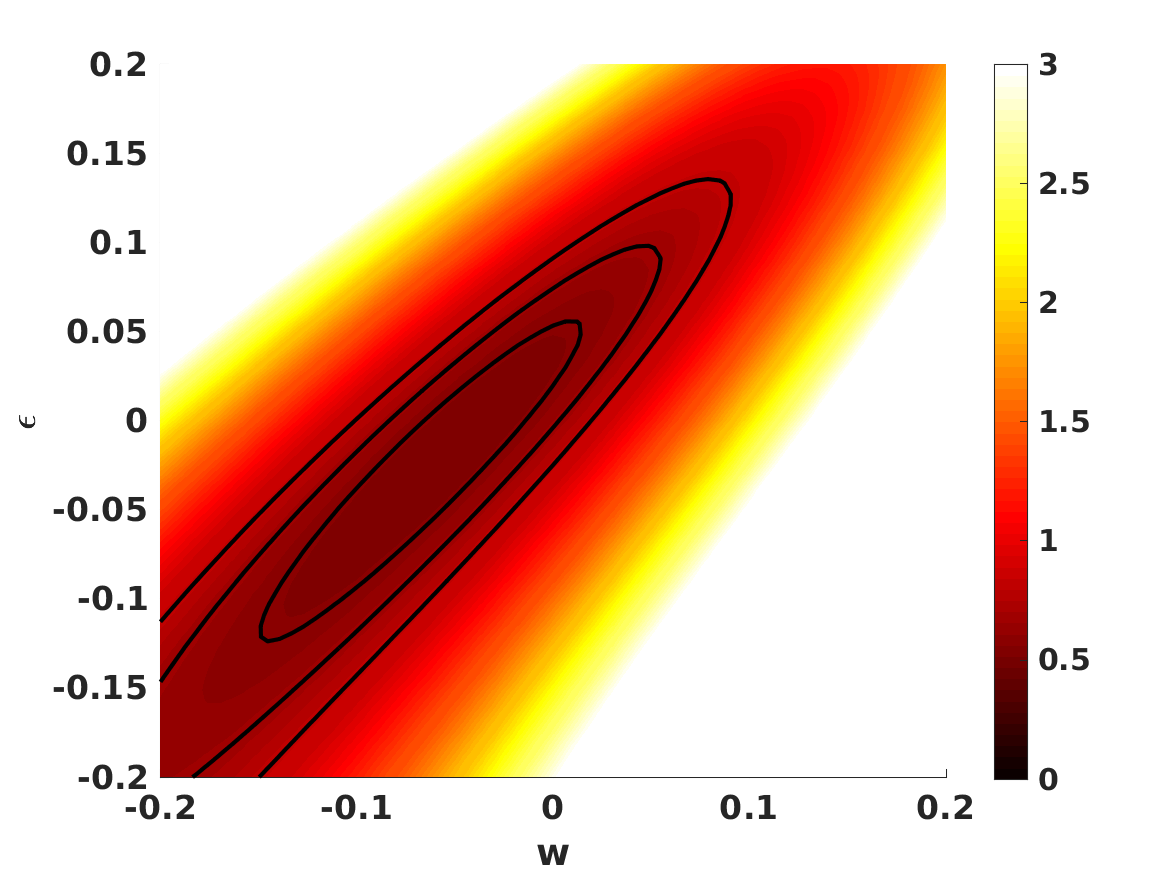}
\includegraphics[width=3.2in,keepaspectratio]{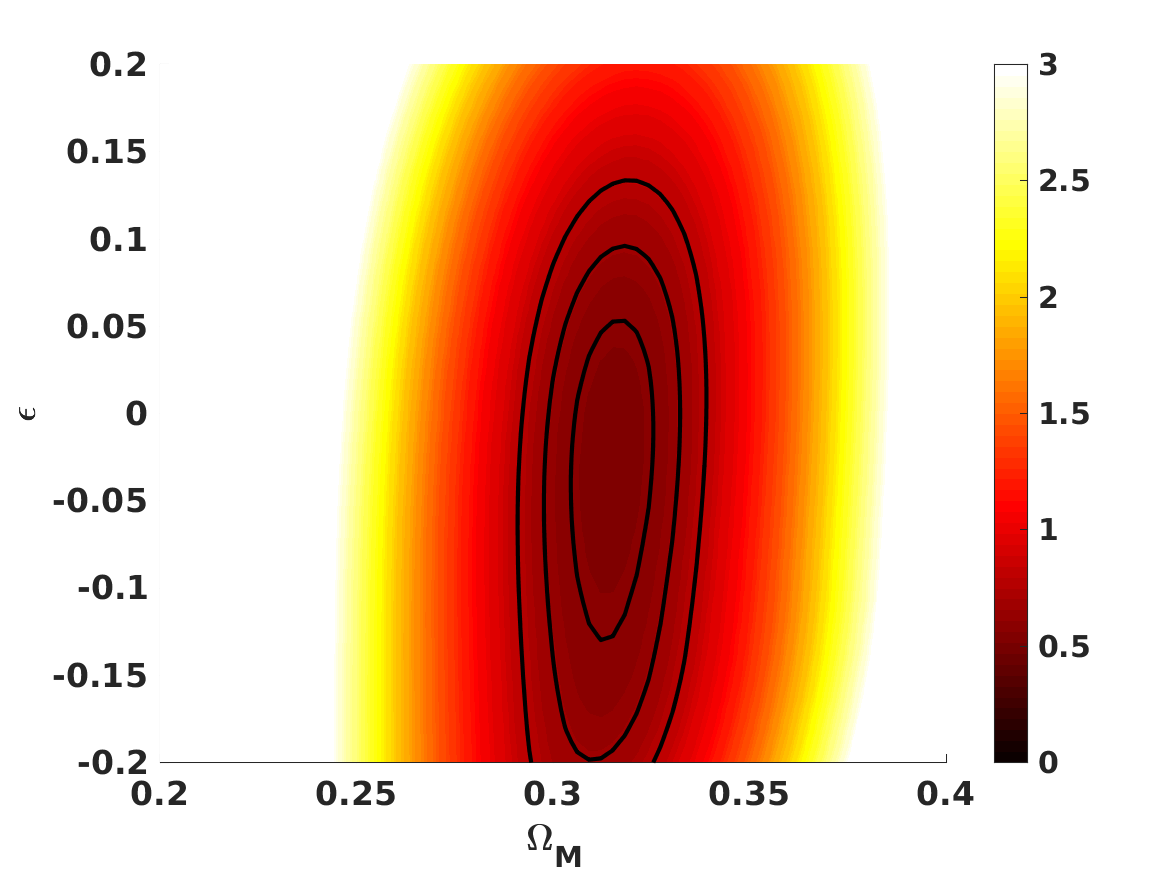}
\includegraphics[width=3.2in,keepaspectratio]{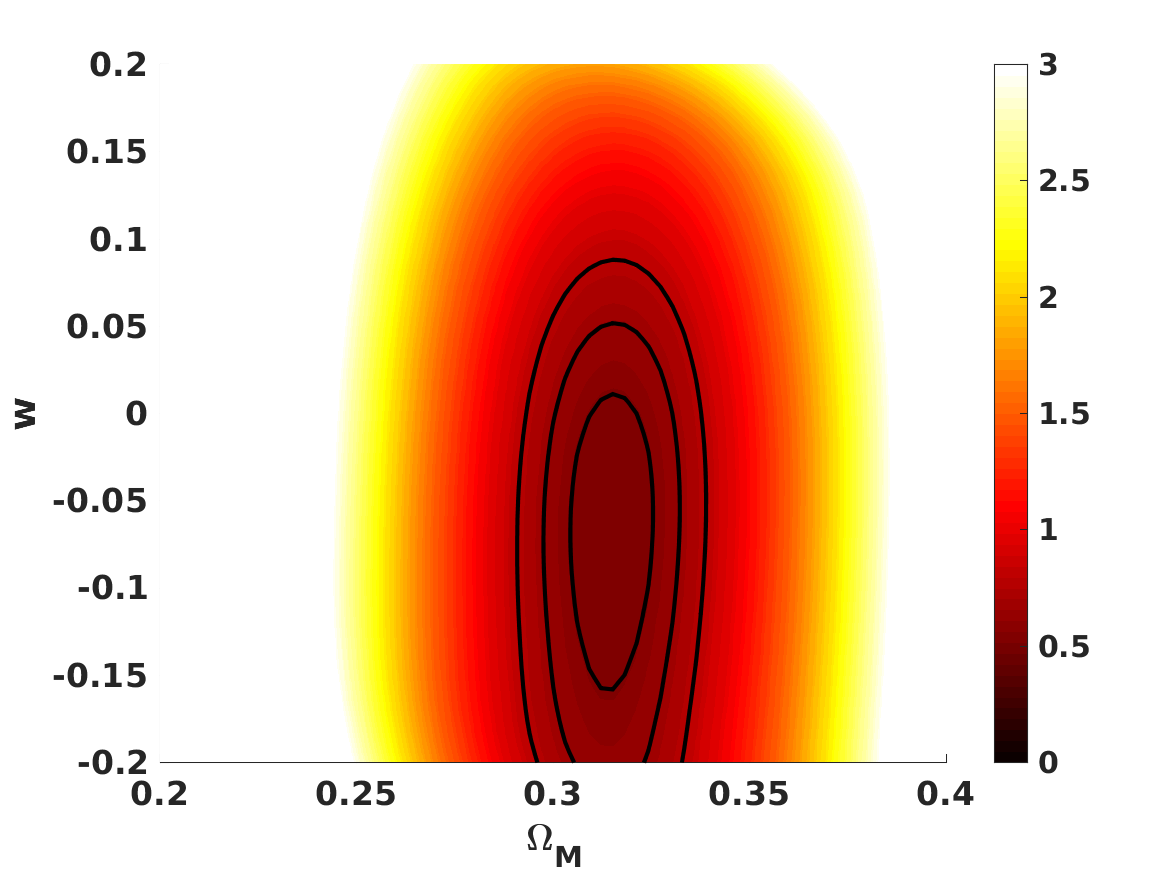}
\end{center}
\caption{\label{fig5}Likelihood constraints in the various 2D planes of the $\epsilon$--$w$--$\Omega_m$ parameter space. The black lines represent the one, two and three sigma confidence levels, and the colormap depicts the reduced chi-square of the fit.}
\end{figure}

The results of the general analysis in this case are shown in Figure \ref{fig5}, which can be compared to the right-hand side panels of Figure \ref{fig3}. The one-sigma posteriors for the two parameters are now
\be
\epsilon=-0.02^{+0.05}_{-0.07}\,
\ee
\be
w=-0.06^{+0.05}_{-0.06}\,.
\ee
In this case the posterior likelihoods are somewhat more asymmetric due to a stronger correlation between $\epsilon$ and the other free parameters in the model, but other than this the results are the expected ones: one-sigma uncertainties and the best-fit value for $\epsilon$ differ from those on $\lambda$ by a factor of about two, while the best-fit value of $w$ is relatively unaffected. And while both best-fit values are slightly negative, there is no statistically significant preference for the presence of torsion.

\section{Conclusions}
\label{sect6}

We have studied a phenomenological but physically motivated class of models for the late-time evolution of the universe allowing for the presence of spacetime torsion, and used low-redshift data (occasionally complemented by a Planck prior on the matter density) to constrain them. This analysis allows us to quantify by how much torsion may contribute to the recent universe, and is also useful as a stress test of $\Lambda$CDM in the sense that it allows us to further study how dark energy may differ from a cosmological constant at low redshifts.

In our analysis we have restricted ourselves to background quantities, thus obtaining a preliminary assessment of the observational viability of these models. An analysis of linear perturbations in this model (necessary, for example, to obtain detailed cosmic microwave background constraints) remains to be done. It is clear from our results that the constraints benefit from a matter prior, for which we have relied on the Planck data \cite{Planck}. We note that it is not {\it a priori} obvious that this prior is applicable to these models, especially if the models are interpreted as fundamental alternatives to $\Lambda$CDM (i.e., if $\Omega_\Lambda=0$  and torsion itself provided the acceleration). However, such a scenario is strongly disfavoured even with low redshift data alone, so this is a moot point. Thus the remaining cosmological niche for these models is a phenomenological one, in which the models are one-parameter extensions of $\Lambda$CDM, with the extra parameter (here denoted $\lambda$) expected to be small. In this case, and although this needs to ultimately be confirmed by a full CMB analysis, we would expect that it is reasonable to use the prior.

Our constraints on $\lambda$ are comparable to (or slightly stronger than) the ones obtained in \cite{Torsion1} from big bang nucleosynthesis, and also agree, at least qualitatively, with those of \cite{Pereira} (which uses different assumptions on the parametrization of torsion and on cosmological model parameters). We have also quantified how these constraints can be improved by analogous low-generation data, and checked the robustness of the steady-state torsion assumption by also studying an alternative parametrization.

In conclusion, we find no statistically significant preference for the presence of torsion. By itself torsion can't be responsible for the acceleration of the universe, and even if taken as an extension of the canonical $Lambda$CDM paradigm the overall contribution to the Universe's energy budget is constrained to be no larger than a few percent (the exact number depending on the underlying assumptions). We also note that our constraints should be seen as conservative. Our analysis has focused on low-redshift background cosmology data, except for the occasional inclusion of a Planck-inspired prior on the matter density. An analysis including a full treatment of the cosmic microwave background should lead to stronger constraints. We leave this interesting analysis for subsequent work.

\section*{Acknowledgements}

We are grateful to Ana Catarina Leite for helpful discussions on the subject of this work. This work was financed by FEDER---Fundo Europeu de Desenvolvimento Regional funds through the COMPETE 2020---Operacional Programme for Competitiveness and Internationalisation (POCI), and by Portuguese funds through FCT---Funda\c c\~ao para a Ci\^encia e a Tecnologia in the framework of the project POCI-01-0145-FEDER-028987.

\bibliographystyle{model1-num-names}
\bibliography{torsion}
\end{document}